\documentclass[aps,prc,twocolumn]{revtex4}
\usepackage{graphics,amssymb}

\begin{document}

\title{Comment on ``Experimental realization of Wheeler's
  delayed-choice GedankenExperiment''}
\author{Travis Norsen}
\affiliation{Marlboro College \\ Marlboro, VT  05344 \\ norsen@marlboro.edu}

\date{Nov. 2, 2006}

\begin{abstract}
A shortcoming in the authors' interpretation of this beautiful new
experiment is pointed out and briefly discussed.
\end{abstract}

\maketitle

The new experimental realization \cite{er}
of Wheeler's delayed-choice thought
experiment by Jacques, Wu, Grosshans, Treussart, Grangier, Aspect, 
and Roch (hereafter, ``the experimenters'') 
is a fantastic achievement.  In the experiment, single 
photons are split at an initial beamsplitter, with the two ``parts''
then propagating along separate paths toward a detection area where
a second beamsplitter can, at the last possible moment, either be
inserted (causing the two ``parts'' to recombine and interfere) or
removed (in which case one may simply observe which of the two paths
was taken by the photon).  

Of course, it is the aspect of delayed-choice which makes this
so puzzling.  With the second beam splitter in place, the 
observed interference can only be
understood if something ``split in half'' and took both paths 
through the interferometer.  But with the second beam splitter
removed, the photon is (with high precision) observed in one or the
other of the two beams exclusively, but never both.  

As the experimenters explain it, ``the striking feature is that the
phenomenon of interference, interpreted as a wave following
simultaneously two paths, is incompatible with our common sense
representation of a particle which implies to follow one route or the
other but not both.'' \cite{er}  But, because
the ``choice'' (made by a Quantum Random Number Generator in this
experimental realization) of whether the second beam splitter is to
be inserted or removed is made \emph{after} the photon has long since
passed the initial beam splitter (at which it presumably would have
to decide whether to split in half and take both paths, or select a
single path) there appears to be a kind of non-local or
backwards-in-time causation.  

Actually, perhaps because he rejected as absurd any such non-local or
reverse-temporal causation, Wheeler himself interpreted the
significance of the thought experiment this way:
\begin{quote}
``Then let the general lesson of this apparent time
inversion be drawn:  `No phenomenon is a phenomenon until it is an
observed phenomenon.'  In other words, it is not a paradox that we
choose what \emph{shall} have happened after `it has \emph{already}
happened.'  It has not really happened, it is not a phenomenon, until
it is an observed phenomenon.''  \cite{wheeler}
\end{quote}
The experimenters are apparently less comfortable with this radically
subjectivist and anti-realist philosophy, and simply claim that the 
experiment demonstrates a surprising sort of causality:
\begin{quote}
``Our realization of Wheeler's delayed-choice GedankenExperiment
demonstrates beyond any doubt that the behavior of the photon in the
interferometer depends on the choice of the observable which is
measured, even when that choice is made at a position and a time such
that it is separated from the entrance of the photon in the
interferometer by a space-like interval.'' \cite{er}
\end{quote}

\begin{center}
*  *  *
\end{center}

But does the experimenters' experiment really establish such 
non-local causation (or, for that mattter, Wheeler's subjectivism)
\emph{``beyond any doubt''}?  

The answer is demonstrably negative.  For a theory exists which can
account for the observed results of Wheeler's delayed-choice experiment
in a completely ordinary, local, common-sensical fashion.   To see
how this is possible, it is helpful to note an additional premise that 
Wheeler and the experimenters use in deducing from the observed
results their respective conclusions.  The premise is this:  each
``individual photon'' is fundamentally, unanalyzably, ontologically
\emph{one thing}.  It is only in the presence of this tacit premise
that the claims
\begin{itemize}
\item[(i)] something took exclusively one of the two available paths
  through the interferometer,  and
\item[(ii)] something took simultaneously both paths through the
  interferometer 
\end{itemize}
form together a logical contradiction which must be avoided by saying
(naive appearances to the contrary notwithstanding) that, really,
only one of (i) and (ii) is true.  And it is precisely saying this 
which implies non-local
causation since \emph{which thing happened} is apparently influenced 
by our (later) choice to insert (or not) the beamsplitter.  

But suppose, as postulated by the pilot-wave theory of de Broglie and
Bohm, that each ``individual photon'' consists of two ontologically
distinct aspects:  a wave \emph{and} a particle.  \cite{bm}
According to this 
theory, which is empirically equivalent to standard quantum theory,
the photon \emph{particle} obeys (i), i.e., it follows a definite
trajectory and thus takes exclusively one or the other of the two
possible paths through the interferometer.  Meanwhile, the
\emph{wave}, in accordance with (ii), takes both paths.  
The trajectory of the particle is influenced by the wave in a way 
that explains exactly why the particle ends up where it ends up and
with precisely the observed empirical frequencies under the various
experimental conditions.  And, crucially, the theory 
does this without in any way
requiring us to posit a spooky backwards-in-time causation (or worse,
dropping altogether the idea that something actually happened between
the production and detection of the photon).  

The theory of de Broglie and Bohm really exists, and really works.
And it provides a stark counterexample to the claim that the
results of Wheeler's delayed-choice experiment \emph{require} ``beyond 
any doubt that the [earlier] behavior of the photon in the
interferometer depends on the [later] choice of the observable which is
measured''.  

It is frustrating that this needs to be pointed out.  The de Broglie -
Bohm theory has existed for more than 50 years.  Moreover, 25 years 
ago, J.S. Bell wrote an entire paper aimed at making this same point
-- that the pilot-wave theory provides an elegant alternative to the
kinds of inferences made from Wheeler's delayed-choice experiment
by physicists who are unduly in the
grip of the orthodox quantum philosophy. \cite{bell}

One can do no better than simply quote Bell's penetrating summary of
Wheeler's argument, and his explanation of how the de Broglie - Bohm
theory eludes Wheeler's conclusion.  First:
\begin{quote}
The decision, to interpose the [beam splitter] or not, is made only
\emph{after} the pulse has passed the slits.  As a result of this
choice the particle \emph{either} falls on one of the two counters,
indicating passage through one of the two [arms of the
interferometer], \emph{or} contributes [to the building of an]
interference pattern after many repititions.  Sometimes the
interference pattern is held to imply `passage of the particle through
both slits' -- in some sense.  Here it seems possible to
\emph{choose}, \emph{later}, whether the particle, \emph{earlier},
passed through one [arm] or two!  Perhaps it is better not to think
about it.  `No phenomenon is a phenomenon until it is an observed
phenomenon.''' \cite{bell}
\end{quote}
Second:  as Bell explains, in  the de Broglie - Bohm theory
\begin{quote}
``the wave always goes  through both [arms] (as is the nature of
waves) and the particle goes through only one (as is the nature of
particles).  But the particle is guided by the wave toward places
where $|\psi|^2$ is large, and away from places where $|\psi|^2$ is
small.  And so if the [second beam splitter] is in position the
particle contributes a spot to the interference pattern ... or if the
plate is absent the particle proceeds to one of the counters.  
\emph{In
neither case is the earlier motion, of either particle or wave,
affeted by the later insertion or noninsertion of the [beam
splitter].}'' (emphasis added) \cite{bell}
\end{quote}

\begin{center}
*  *  *
\end{center}

There is a certain irony here  associated with 
the fact that most physicists (at least, among those who have
even heard of it) reject the de Broglie - Bohm theory because it is
explicitly non-local.  It's certainly correct that it is:  
the theory posits a
mechanism whereby goings-on at the location of one particle, can
affect the trajectory of another, distant (entangled) particle,
sooner than signals propagating at the speed of light would permit.
And this non-locality is crucial to the theory's ability to match
the empirically correct predictions of standard quantum theory.

But the rejection of the pilot-wave theory on this basis is 
fallacious, for, as proved by Bell's Theorem, \emph{any} theory which
is in agreement with the experimental tests of Bell's Inequality must
display a similar non-local causality.  

Proponents of orthodox quantum theory, however, are often confused
about this and think of \emph{their} theory as perfectly local.  But,
simply put, it isn't:  either one accepts (with Wheeler) an 
anti-realism which prevents the theory from saying anything dynamical
at all (about, for example, photons), such that it simply
doesn't say anything about the kinds of processes to which the terms
``local'' and ``non-local'' apply; or (like the
experimenters) one must admit that the dynamics of the theory
(in particular processes involving ``measurement'') are manifestly
non-local.   Either one stubbornly insists that the theory doesn't
say \emph{anything}, or one admits that what it says involves 
non-locality.  The point is, in neither case can one claim that 
the theory provides a local description of the dynamics of photons
(etc.). 

The primary insight offered by the Wheeler delayed-choice experiment 
is that, while (as proved by Bell) any theory which agrees with
\emph{all} of the quantum mechanical predictions must be non-local,
\emph{some} theories display that troubling non-locality more often
or more blatantly than others.  
Here is a situation which can be explained simply and locally by the
de Broglie - Bohm theory, but whose explanation in terms of the 
orthodox quantum theory requires non-locality or worse.
And so the irony is that those who reject the de
Broglie - Bohm theory because it is non-local, and favor instead the
standard version of quantum theory, unwittingly end up favoring 
something that is (in the sense just elaborated) \emph{more non-local}
than the theory they reject because it is non-local.  
Co-opting (for a purpose he wouldn't like) 
an infamous passage of N.D. Mermin: those for whom
non-locality is anathema should (in response to Wheeler's experiment) 
reject orthodox quantum theory and
flock to the pilot-wave picture!  \cite{mermin}

Of course, the real lesson here is just that anyone not conversant
with the pilot-wave theory is severely hampered when it comes to 
interpreting the significance and meaning of fundamental experiments
in physics.  As Bell noted,
\begin{quote}
``Even now the de Broglie - Bohm picture is generally
ignored, and not taught to students.  I think this is a great loss.
For that picture exercises the mind in a very salutary way.'' \cite{bell2}
\end{quote}
And so one is naturally led to wonder, again following Bell:
\begin{quote}  
``Why is the pilot wave picture ignored in textbooks?  Should
it not be taught, not as the only way, but as an antidote to the
prevailing complacency?  To show that vagueness, subjectivity, and
indeterminism are not forced on us by experimental facts, but by
deliberate theoretical choice?''  \cite{bell3}
\end{quote}
Tragically, decades later, physicists (who apparently still remain
ignorant of the important lessons of de Broglie and Bohm) are still
making the latter choice (apparently without even realizing they 
are making a choice).  One can only hope that the more reasonable
choice -- of acknowledging the real existence of the pilot 
wave theory and learning the important lessons it has to teach -- 
will be not much longer delayed.

\end{document}